\documentclass[prd,showpacs,preprintnumbers,
a4paper, nofootinbib ,twocolumn,
]{revtex4}
  
\usepackage{amsmath,amssymb,axodraw}

\usepackage{epsfig}  
  
\usepackage{graphicx}               
\def\K{\kern.25em}
\newdimen\diaght
\def\Ln{%
  \ifdim \diaght<-6pt
    \let\Ln=\relax
  \else \advance\diaght by-.3pt
    \raise \diaght \hbox{\vrule height .3pt depth .3pt width .3pt}%
  \fi \Ln}  
\newcommand{\nc}{\newcommand}  

\nc{\beq}{\begin{equation}}  
\nc{\eeq}{\end{equation}}  
\nc{\beqa}{\begin{eqnarray}}  
\nc{\eeqa}{\end{eqnarray}}  
\nc{\bea}{\begin{eqnarray}}  
\nc{\eea}{\end{eqnarray}}  
\nc{\ra}{\rightarrow}  
\nc{\lsim}{\begin{array}{c}\,\sim\vspace{-21pt}\\< \end{array}}  
\nc{\gsim}{\begin{array}{c}\sim\vspace{-21pt}\\> \end{array}}

\nc{\LL}{L}  
\nc{\vv}{\tilde{v}}  
\nc{\GG}{\widetilde{G}}  
\nc{\ktilde}{\tilde{k}}  
\nc{\MKK}{\ensuremath{M_{\rm KK}}}
\nc{\gc}{\ensuremath{G_{\rm c}}}
\nc{\g}[1]{\ensuremath{g_{\rm #1}}}
\nc{\ZZ}{\ensuremath{\mathcal{{\cal{Z}}}}}

\newcommand{\GeV}{\mbox{GeV}}

\newcommand{ \slashchar }[1]{\setbox0=\hbox{$#1$}   
   \dimen0=\wd0                                     
   \setbox1=\hbox{/} \dimen1=\wd1                   
   \ifdim\dimen0>\dimen1                            
      \rlap{\hbox to \dimen0{\hfil/\hfil}}          
      #1                                            
   \else                                            
      \rlap{\hbox to \dimen1{\hfil$#1$\hfil}}       
      /                                             
   \fi}   
  
\begin{document}
\preprint{FERMILAB-PUB-08-504-T}

\title{Unified dark matter model in a singlet extension of the universal
   extra dimension model}

\author{Yang Bai${}^{a}$ \thanks{email: bai@fnal.gov}}
\author{Zhenyu Han${}^{b}$ \thanks{email:zhenyuhan@physics.ucdavis.edu}}
\address{  \vspace{3mm}
${}^{a}$Theoretical Physics Department, Fermilab, Batavia, Illinois 60510,
${}^{b}$Department of Physics, University of California, Davis, CA 95616}


\pacs{11.25.Mj, 95.35.+d}
 
\begin{abstract} 
We propose a dark matter model with standard model singlet extension
of the universal extra dimension model (sUED) to explain the recent
observations of ATIC, PPB-BETS, PAMELA and DAMA. Other than the
standard model fields propagating in the bulk of a 5-dimensional
space, one fermion field and one scalar field are introduced and both
are standard model singlets. The zero mode of the new fermion is
identified as the right-handed neutrino, while its first KK mode is
the lightest KK-odd particle and the dark matter candidate.  The
cosmic ray spectra from ATIC and PPB-BETS determine the dark matter
particle mass and hence the fifth dimension compactification scale to
be 1.0--1.6~TeV. The zero mode of the singlet scalar field with a mass
below 1~GeV provides an attractive force between dark matter
particles, which allows a Sommerfeld enhancement to boost the
annihilation cross section in the Galactic halo to explain the PAMELA
data. The DAMA annual modulation results are explained by coupling the
same scalar field to the electron via a higher-dimensional
operator. We analyze the model parameter space that can satisfy the
dark matter relic abundance and accommodate all the dark matter
detection experiments. We also consider constraints from the diffuse
extragalactic gamma-ray background, which can be satisfied if the dark
matter particle and the first KK-mode of the scalar field have highly
degenerate masses.  
\end{abstract}    
\maketitle

\section{Introduction} 
Recently, there have been many pieces of evidence on detection of dark
matter (DM) from either  direct searches or  indirect searches.  The
DAMA/LIBRA collaboration has published their new results and confirmed
and reinforced their detection of an annual modulation in their signal
rate. Combined with the DAMA/NaI data, they have interpreted this
observation as evidence for dark matter particles at the $8.2\sigma$
confidence level~\cite{Bernabei:2008yi}.  The ATIC-2 experiment has
reported an excess in its preliminary $e^{+}\,+\,e^{-}$ data at
energies of $500-800$~GeV~\cite{Chang:2005}. This is confirmed
recently by the PPB-BETS balloon experiment~\cite{Torii:2008xu}. It
can be naturally explained by dark matter annihilation into electrons
and positrons. The PAMELA collaboration has published their results
showing an anomalous increase of the positron fraction in the
energy range of $10-100$~GeV~\cite{Adriani:2008zr}. All of these
experimental results are possible indications of detection of dark
matter. In this paper, we propose to explain all these experimental
results in terms of a concrete particle physics model.  

There are well motivated models containing unbroken discrete
symmetries and providing dark matter candidates in the literature. In
the extensively studied Minimal Supersymmetric Standard Model (MSSM), the $R$-parity protects the lightest supersymmetric particle from decay~\cite{Jungman:1995df}. In the
Universal Extra Dimension model (UED)~\cite{Appelquist:2000nn}, the
Kaluza Klein (KK)-parity keeps the lightest KK-odd particle stable and therefore also
provides a dark matter candidate~\cite{Servant:2002aq}. In this paper,
we focus on the explanation of dark matter experiments based on the UED
model. However, in the minimal UED model, the dark matter candidate,
KK-photon, is difficult to account for the PAMELA results because of
their small annihilation rate to electrons and positrons in the
Galactic halo. Recent studies have shown that the PAMELA results can
be explained if the dark matter particles mainly annihilate into
electron and positron pairs, and there exists a large boost factor to
increase the annihilation cross section in the Galactic
halo~\cite{Cirelli:2008jk}. The large boost factor can be obtained
through the Sommerfeld enhancement effect if there is a new long-range
force attractive between two dark matter
particles~\cite{ArkaniHamed:2008qn} (see also~\cite{models} for other
particle physics models that explain the PAMELA results). If the
particle mediating the long-range force only couples to electrons,
then the elastic scattering of dark matter on the electron may explain
the DAMA results without contradicting the null results from other
direct dark matter experiments like CDMS~\cite{CDMS} and
XENON~\cite{XENON}.   

Hence, additional ingredients are needed in the UED model. In this
paper, we  explore this possibility by studying the standard model
singlet extension of the UED model (sUED). Other than the standard
model (SM) fields propagating in the 5-dimensional bulk, we introduce
two new SM singlet fields: one fermion field and one scalar field. The
zero mode of the fermion field $\nu_{R}$ plays the role of the
right-handed neutrino and generates the neutrino Majorana mass through the
see-saw mechanism. The first KK-mode of the fermion field contains
two Weyl fermions with the lightest one called $\chi_{-}$ as the
lightest KK-odd particle. To simplify our discussions, we only
include one generation of right-handed neutrinos, while the
generalization to more generations is staightforward. The zero mode
of the scalar field $s_{0}$ is chosen to have a mass below 1 GeV,
which couples to the right-handed neutrino field 
through a renormalizable operator and hence provides a long-range
force for the dark matter candidate $\chi_{-}$. In order to  explain
the PAMELA results, we also couple the light scalar field $s_{0}$ to
the electron field through a higher-dimensional operator and let
$s_{0}$ mainly decay into a pair of electrons. The same coupling of
$s_{0}$ to the electron field can also generate a large elastic
scattering cross section betweeen $\chi_{-}$ and the electron to
explain the DAMA results. 

The paper is organized as follows. In section~\ref{sec:model}, we
present the field content and the Lagrangian of our model. We
analyze the particle spectrum and identify $\chi_{-}$ as the
lightest KK-odd particle and the dark matter candidate. In
section~\ref{sec:relic}, we calculate the relic abundance of the dark
matter candidate. We also take the co-annihilation effects into
account in this section. We illustrate how to accommodate the ATIC-2,
PPB-BETS and PAMELA results, and the DAMA results in
section~\ref{sec:pamela} and \ref{sec:dama} respectively. In
section~\ref{sec:diffuse}, we show how to evade the constraints from 
diffuse extragalactic gamma-ray background today by making two KK-odd
particles highly degenerate and lowering the kinetic decoupling
temperature of the dark matter. Finally, we conclude in
section~\ref{sec:conclusion}.

\section{The Model}
\label{sec:model}
We consider all SM fields, a new SM singlet fermion field $N$ and a
new SM singlet scalar field $S$, propagating in one extra dimension,
which is compactified on an $S^{1}/Z_{2}$ orbifold with the
fundamental region $0\le y \le \pi\, R$.  We have the action of our
model as follows 
\begin{widetext}
\beqa
S_{5D}&=&\int d^{4}\,x\,\int^{\pi\,R}_{0} d\,y\left[
{\cal L}_{SM}\,-\,\sqrt{\pi
  R}\,y_{\nu}\,\bar{L}\,\tilde{H}\,N\,-\,\frac{1}{2}\,m\,N^{T}\,{\cal
  C}_{5}\,N\,-\,\frac{1}{2}\,\mu^{2}\,S^{2} \right.  
 \nonumber \\
&&\left. \,-\,(\pi R)^{2}\,y^{\prime}_{e}\,S\,\bar{L}\,H\,E\,-\,(\pi
 R)^{2}\,y^{\prime}_{_D}\,S\,\bar{L}\,\tilde{H}\,N\,-\,\frac{1}{2}\,\sqrt{\pi
   R}\,y_{_M}\,S\,N^{T}\,{\cal C}_{5}\,N\,+\,h.c.\, 
\right]\,. 
\eeqa 
\end{widetext}
Here, $y_{\nu}$ and $y^{\prime}_{i}$ are dimensionless parameters; $L$
and $H$ are $SU(2)$ doublets and give us the four-dimensional field
$\ell_{L}=(\nu_{L}, e_{L})^{T}$ and the Higgs doublet $h$;  $E$ is an
$SU(2)$ singlet and corresponds to four-dimensional field $e_R$;
$\tilde{H}\equiv i\,\sigma_2\,H^*\,$ with $\sigma_2$ as the second
Pauli matrix.  $N$ contains the right-handed neutrino $\nu_{R}$ as its
zero mode; ${\cal C}_{5}$ is the 5-d charge-conjugate operator, ${\cal
  C}_{5}\equiv
i\,\gamma_{0}\gamma_{2}\gamma_{5}$~\cite{Pilaftsis:1999jk}. In our
analysis, we will neglect the family indices, but note that it is easy
to extend our model to include more generations of fermions.  

In the following analysis, we will choose $1/R=O({\rm TeV})$ and $m
\sim \mu = O({\rm GeV})$ to explain the recent observations. We also
note that the choice of a positive mass for the scalar field is not
mandatory. An alternative approach is to consider a potential for $S$ with a
negative mass such that the zero-mode of $S$ develops a vacuum
expectation value, which can replace the parameter $m$ through the
Yukawa coupling $y_M$. Since there is a hierarchy between the scales
$\mu$ and $1/R$, the mass of the zero-mode $S$ of $O({\rm GeV})$ is
not stable against radiative corrections, in a similar way to the
Higgs mass in the standard model. A more sophisticated process for building the model
is required to address this problem.

The scalar field, $S$, is decomposed to 4-d fields as
\beq
S(x^{\mu}, y)\,=\,\frac{1}{\sqrt{\pi\,R}}\left[
s_{0}(x^{\mu})\,+\,\sqrt{2}\,\sum_{j\ge 1}s_{j}(x^{\mu})\cos{(\frac{j\,y}{R})}
\right]\,.
\eeq
The 5-d spinor field $N\equiv  (\xi , \bar{\eta})^{T}$, with
$\bar{\eta}\equiv i\,\sigma_{2}\,\eta^{*}$. We choose the
Neumann-Neumann boundary condition for $\xi$ and hence the
Dirichlet-Dirichlet boundary condition for $\eta$. $\nu_{R}$ is the
zero mode of $\xi$. The fields $\xi$ and $\eta$ are decomposed as 
\beqa
\xi (x^{\mu}, y)&=&\frac{1}{\sqrt{\pi\,R}}\left[
\nu_{R}(x^{\mu})\,+\,\sqrt{2}\,\sum_{j\ge 1}\,[\xi_{j}(x^{\mu})\cos{(\frac{j\,y}{R})} ]
\right]\,, \nonumber
\\
\eta (x^{\mu}, y)&=&\sqrt{\frac{2}{\pi\,R}}\,\sum_{j\ge 1}\,[\eta_{j}(x^{\mu})\sin{(\frac{j\,y}{R})}  ]\,.
\eeqa
After integrating out the fifth dimension and after breaking the eletroweak
symmetry, we arrive at the following 4-d effective Lagrangian 
\begin{widetext}
\beqa
-{\cal
  L}_{4d}&=&y_{\nu}\,v\,\bar{\nu}_{L}\,\nu_{R}\,+\,\frac{1}{2}\,m\,\nu^{T}_{R}\,i\sigma_{2}\,\nu_{R}\,+\,\frac{1}{2}\,\mu^{2}\,s_{0}^{2}\,+\,y_{e}\,s_{0}\,\bar{e}_{L}\,e_{R}\,+\,y_{_D}\,s_{0}\,\bar{\nu}_{L}\,\nu_{R}\,+\frac{1}{2}\,y_{_M}\,s_{0}\,\nu^{T}_{R}\,i\sigma_{2}\,\nu_{R}  
\nonumber \\ 
&& +\,\frac{1}{2}\,(\mu^{2}\,+\,\frac{1}{R^{2}})\,s_{1}^{2}\,+\,\frac{1}{2}\,m\,(\xi^{T}_{1}\,i\sigma_{2}\,\xi_{1}\,+\,\eta^{T}_{1}\,i\sigma_{2}\,\eta_{1})\,+\,\frac{1}{R}\,\eta^{T}_{1}\,i\sigma_{2}\,\xi_{1} \nonumber  \\
&&+\,y_{_D}\,s_{1}\,\bar{\nu}_{L}\,\xi_{1}\,+\,\frac{1}{2}\,y_{_M}\,s_{0}\,(\xi^{T}_{1}\,i\sigma_{2}\,\xi_{1}\,+\,\eta^{T}_{1}\,i\sigma_{2}\,\eta_{1})  
\,+\,\,y_{_M}\,s_{1}\,\xi^{T}_{1}\,i\sigma_{2}\,\nu_{R}\,+\,h.c.\,+\,\cdots\,.
\eeqa
\end{widetext}
Here we only keep  the zeroth and first KK modes of the particles in
the Lagrangian; $v=174$~GeV is the vacuum expectation value of the
Higgs field; $y_{e}\equiv y_{e}^{\prime}\,\pi\,v\,R$ and $y_{_D}\equiv
y_{_D}^{\prime}\,\pi\,v\,R$. In the mass eigenbasis, $\chi_{-}\equiv
i\, (\xi_{1} -\eta_{1})/\sqrt{2}$ and $\chi_{+}\equiv  (\xi_{1}
+\eta_{1})/\sqrt{2}$, we have 
\begin{widetext}
\beqa
-{\cal L}_{4d}&=&\frac{1}{2}\,m_{\nu}\,\nu^{T}_{L}\,i\sigma_{2}\,\nu_{L}\,+\,\frac{1}{2}\,m\,\nu^{T}_{R}\,i\sigma_{2}\,\nu_{R}\,+\,\frac{1}{2}\,\mu^{2}\,s_{0}^{2}\,+\,y_{e}\,s_{0}\,\bar{e}_{L}\,e_{R}\,+\,y_{_D}\,s_{0}\,\bar{\nu}_{L}\,\nu_{R}  \,+\,\frac{1}{2}\,y_{_M}\,s_{0}\,\nu^{T}_{R}\,i\sigma_{2}\,\nu_{R}
\nonumber \\
&&
\,+\,\frac{1}{2}\,M^{2}_{s}\,s_{1}^{2}\,+\,\frac{1}{2}\,M_{+}\,\chi^{T}_{+}\,i\sigma_{2}\,\chi_{+}\,+\,\frac{1}{2}\,M_{-}\,\chi^{T}_{-}\,i\sigma_{2}\,\chi_{-}
\,+\,\frac{y_{_D}}{\sqrt{2}}\,s_{1}\,\bar{\nu}_{L}\,\chi_{+} \,-\,\frac{i\,y_{_D}}{\sqrt{2}}\,s_{1}\,\bar{\nu}_{L}\,\chi_{-}
  \nonumber \\
&&\,+\,\frac{1}{2}\,y_{_M}\,s_{0}\,(\chi^{T}_{+}\,i\sigma_{2}\,\chi_{+}\,-\,\chi^{T}_{-}\,i\sigma_{2}\,\chi_{-}) \,+\,\frac{y_{_M}}{\sqrt{2}}\,s_{1}\,\chi^{T}_{+}\,i\sigma_{2}\,\nu_{R}\,-\,\frac{i\,y_{_M}}{\sqrt{2}}\,s_{1}\,\chi^{T}_{-}\,i\sigma_{2}\,\nu_{R}\,+\,h.c.\,+\,\cdots\,.
\label{Eq:lag}
\eeqa
\end{widetext}
Here $m_{\nu}=y_{\nu}^{2}v^{2}/m$ is the left-handed neutrino mass
through the see-saw mechanism (we will choose the energy scale for $m$
to be around one GeV, so $y_{\nu}$ needs to be very small to fit the
neutrino mass. For $y_\nu\approx 6\times 10^{-8}$, we have
$m_{\nu}\sim 0.1$~eV. Although this neutrino Yukawa coupling is three
order of magnitude smaller than the electron Yukawa coupling, the
see-saw mechanism is still playing a role here, otherwise the neutrino
Yukawa coupling needs to be $\sim 10^{-12}$); the right-handed
neutrino mass is approximately $m$ assuming $y_{\nu}v\ll m$;
$M^{2}_{s}=(\mu^{2}+1/R^{2})$, which is the mass of the first KK mode of the
scalar field; $M_{\pm}=1/R\pm m$ which are positive for $m\ll 1/R$;
$2\,m_{e}<\mu<m$, so the right-handed neutrino $\nu_{R}$ can decay to
$\nu_{L}$ plus $s_{0}$, and $s_{0}$ can decay to two electrons. In the
minimal UED model, after taking radiative corrections into account,
all first KK modes have masses above the compactification scale
$1/R$~\cite{Cheng:2002iz}. The fermion Yukawa coupling and the scalar
quartic coupling in general will lower the first KK-mode masses.
Therefore, we anticipate that after radiative corrections, the three
new KK-odd particles $s_{1}$, $\chi_{+}$ and $\chi_{-}$ have masses
below other SM KK modes. Due to theoretical uncertainties including
Brane-localized terms, we will keep their masses as free parameters.  
Furthermore, we assume $M_{+}>M_{s} > M_{-}$, therefore the lightest
KK-odd particle $\chi_{-}$ is the dark matter candidate in this
model. The $\chi_{+}$ field decays into  $s_{1}$ plus $\nu_{L}$, while
$s_{1}$ mainly decays into $\chi_{-}$ plus $\nu_{L}$ when
$M_{s}-M_{-}< m$ (the decay channel of $s_{1}$ to $\chi_{-}$ plus
$\nu_{R}$ is kinematically forbidden). 

\section{Annihilation, Co-annihilation and Relic Abundance}
\label{sec:relic}

The present relic abundance of dark matter is related to the
pair-annihilation rate in the non-relativistic limit by the sum of
the quantities, $a(X) = \langle v\,\sigma \rangle $ with $v\sim 0.3$
to be the relative velocity between the dark matter particles. For
simplicity, we only consider $s$-wave channel annihilation in this
paper because the $p$-wave channel is suppressed by ${\cal
  O}(v^{2})$. The present dark matter abundance from WMAP
collaboration, $0.096 < \Omega\,h^{2}< 0.122\,(2\sigma)$, requires
$a_{\rm tot} = 0.81\pm 0.09\,{\rm
  pb}$~\cite{Birkedal:2004xn}\cite{Bai:2008cf}, assuming the dark 
matter candidate  in our model can make up all the dark matter.  

Since the three lightest KK-odd particles have almost degenerate
masses, we need to consider co-annihilations among these particles.
The effective annihilation cross section~\cite{Griest:1990kh} is  
\beq
\sigma_{\rm
  eff}\,=\,\sum^{3}_{ij}\sigma_{ij}\frac{g_{i}g_{j}}{g^{2}_{\rm eff}}
 (1+\Delta_{i})^{3/2}\,(1+\Delta_{j})^{3/2}\,e^{-x(\Delta_{i}+\Delta_{j})}\,, 
\eeq
with $\sigma_{ij}=\sigma(X_{i}X_{j}\rightarrow {\rm SM\;
  particles})$. Here, $X_{i}$ represents the three lightest particles
in our model with $i=1$ for $\chi_{-}$, $i=2$ for $\chi_{+}$ and $i=3$
for $s_{1}$; $\Delta_{i}\,=\,(M_{i}-M_{-})/M_{-}$; $g_{i}$ is the
number of degrees of freedom of the $i$'s particle: $g_{1,2}=2$ for
$\chi_{\pm}$ and $g_{3}=1$ for $s_{1}$; $g_{\rm eff}$ is defined to be 
\beq
g_{\rm eff}\,=\,\sum^{3}_{i}\,g_{i}\,(1\,+\,\Delta_{i})^{3/2}\,e^{-x\,\Delta_{i}}\,.
\eeq
To simplify our calculation, we will choose $x= x_{\rm F} =
M_{-}/T_{\rm F} \approx 20$ ($T_{\rm F}$ is the dark matter freeze-out
temperature). For nearly degenerate masses such that $\Delta_{i} \ll
1/x$, the exponential part of the above equation approximately equals
to one and is independent of the freeze-out temperature.  When the
three lightest KK-odd particles have nearly degenerate masses or
satisfy $\Delta_{i}< 0.01$, which is the case in our model, we have  
\beqa
\sigma_{\rm eff}&=&\frac{4}{25}(\sigma_{--}\,+\,2\,\sigma_{-+}\,+\,\sigma_{++})
\nonumber \\
&+&\frac{4}{25}(\sigma_{-s}\,+\,\sigma_{+s})\,+\,\frac{1}{25}\sigma_{ss}\,.
\label{eq:sigmaeff1}
\eeqa
Since the operators associated with $y_e$ and $y_{_D}$ are
higher-dimensional operators, it is natural to have $y_{e}\,,
y_{_D}\ll y_{_M}$. In the following, we only keep the largest Yukawa
coupling $y_{M}$ in calculating the annihilation cross section. The
dominant self-annihilation channel of $\chi_{-}$'s  is
$\chi_{-}\,\chi_{-}\,\rightarrow\,\nu_{R}\,\nu_{R}$ in the $t$-channel
by exchanging the $s_{1}$ field (the $s$-channel diagram by exchanging
$s_{0}$ field has zero contribution to the $s$-wave annihilation, and
is neglected here. For the same reason, we also neglect the
annihilation channel $\chi_{-}\,\chi_{-}\rightarrow s_{0}\,s_{0}$). To
leading order in the relative velocity, $v$, of two $\chi_{-}$'s and
neglecting $\nu_{R}$ mass in the limit $m\ll 1/R$, the annihilation
cross section is  
\beq
v\,\sigma_{--}\,=\,\frac{y_{_M}^{4}\,M_{-}^{2}}{64\,\pi\,(M_{-}^{2}\,+\,M_{s}^{2})^{2}}\,+\,{\cal O}(v^{2})\,.
\eeq
The annihilation cross section $\sigma_{++}$ of
$\chi_{+}\,\chi_{+}\,\rightarrow\,\nu_{R}\,\nu_{R}$ has a similar
formula by replacing $M_{-}$ with $M_{+}$. The co-annihilation cross
section of $\chi_{-}\,\chi_{+}\,\rightarrow\,\nu_{R}\,\nu_{R}$ is from
the $t$-channel diagram by exchanging $s_{1}$ and has the formula 
\beq
v\,\sigma_{-+}\,=\,\frac{y_{_M}^{4}\,(M_{-}\,+\,M_{+})^{2}}{256\,\pi\,(M_{-}\,M_{+}\,+\,M_{s}^{2})^{2}}\,+\,{\cal O}(v^{2})\,.
\eeq
The co-annihilation cross section $\sigma_{-s}$ of
$\chi_{-}\,s_{1}\,\rightarrow\,s_{0}\,\nu_{R}$ by exchanging
$\chi_{-}$ in the $t$-channel is calculated to be 
\beq
v\,\sigma_{-s}\,=\,\frac{y_{_M}^{4}\,(M_{-}\,-\,M_{s})^{2}}{64\,\pi\,M_{-}^{2}\,M_{s}\,(M_{-}\,+\,M_{s})}\,+\,{\cal O}(v^{2})\,.
\eeq
A similar formula for $\sigma_{+s}$ can be obtained by changing
$M_{-}$ to $M_{+}$. Finally, for the self-annihilation of $s_{1}$, the
annihilation process is $s_{1}\,s_{1}\,\rightarrow\,\nu_{R}\,\nu_{R}$
by exchanging $\chi_{-}$ and $\chi_{+}$ in the $t$-channel. It has the
following formula 
\beq
v\,\sigma_{ss}\,=\,\frac{y_{_M}^{4}\,(M_{-}\,-\,M_{+})^{2}\,(M_{s}^{2}\,-\,M_{1}\,M_{2})^{2}  }{8\,\pi\,(M_{s}^{2}\,+\,M_{-}^{2})^{2}\,(M_{s}^{2}\,+\,M_{+}^{2})^{2}}\,+\,{\cal O}(v^{2})\,.
\eeq
When $M_{-}$, $M_{+}$ and $M_{s}$ are nearly degenerate, we use the
parameter $M_{-}$ to represent those three variables. From
Eq.~(\ref{eq:sigmaeff1}), we have 
\beq
v\,\sigma_{\rm eff}\,=\,\frac{y_{_M}^{4}}{400\,\pi\,M_{-}^{2}}\,+\,{\cal O}(v^{2})\,.
\label{eq:relic}
\eeq
Therefore, the quantity $a_{\rm tot}$ in our model is 
\beq
a_{\rm tot}\,=\,\frac{y_{_M}^{4}}{400\,\pi\,M_{-}^{2}}\,\approx\,y_{_M}^{4}\,(\frac{1~{\rm TeV}}{M_{-}})^{2}\,\times\,0.32~{\rm pb}\,.
\eeq
For the dark matter mass of 1\,--\,1.6~TeV, we need to choose
$y_{_M}\approx$ 1.2\,--\,1.6  to satisfy the current dark matter relic
abundance.  

\section{ATIC, PPB-BETS and PAMELA}
\label{sec:pamela}

The ATIC-2 balloon experiment reported an excess  in the
$e^{+}\,+\,e^{-}$ energy spectrum between
$500-800$~GeV~\cite{Chang:2005}. This has been confirmed recently by the
PPB-BETS balloon experiment~\cite{Torii:2008xu}. One explanation of
this excess is that the dark matter particles annihilate into
electrons.  

\begin{figure}[ht!]
\centerline{ \hspace*{0.5cm}
\includegraphics[width=0.45\textwidth]{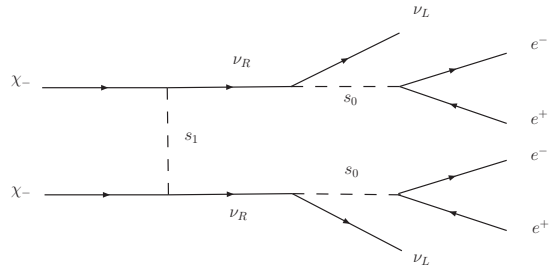}
} 
\caption{Feynman diagram of the main annihilation channel of $\chi_{-}$ in our model.}
\label{fig:relic}
\end{figure}

Specific to our model, the dark matter candidate $\chi_{-}$ mainly
annihilates to the right-handed neutrinos $\nu_{R}$, which subsequently
decay into $\nu_{L}\,+\,s_{0}$. Because the $s_{0}$ has a mass below
$\nu_{R}$ and above twice of the electron mass, it dominantly decays
into two electrons. The process chain is  
\beq
\chi_{-}\,\chi_{-}\,\rightarrow\,\nu_{R}\,\nu_{R}\,\rightarrow\,\nu_{L}\,s_{0}\,\nu_{L}\,s_{0}\,\rightarrow\,\nu_{L}\,e^{+}\,e^{-}\,\nu_{L}\,e^{+}\,e^{-}\,, 
\eeq
and the Feynman diagram is shown in Fig.~\ref{fig:relic}
%
%

\begin{figure}[ht!]
\centerline{ \hspace*{0.1cm}
\includegraphics[width=0.5\textwidth]{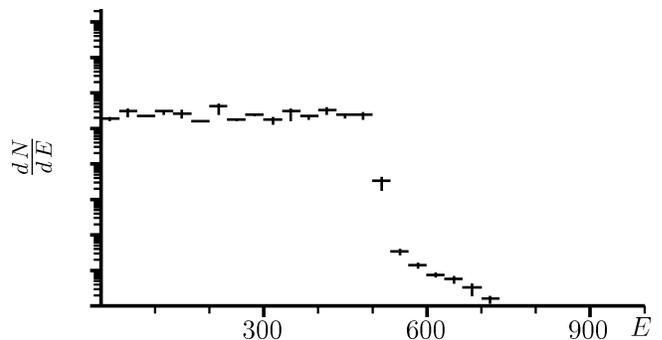}
} 
\caption{The energy density distribution as a function of the positron
  energy $E$  (in GeV) for 1~TeV dark matter mass. The errors on this
  plot come from uncertainties of numerical simulations.} 
\label{fig:density}
\vspace{0.3 cm}
\end{figure}

Neglecting all particles' masses except those of $\chi_{-}$ and
$s_{1}$, each of the four electrons has a nearly flat energy spectrum
with the maximum energy of a half of the dark matter mass
$M_{-}$. This is because each right-handed neutrino $\nu_{R}$ carries
energy of the dark matter mass $M_{-}$. After it decays into a fermion
and a scalar, the scalar field $s_{0}$ carries approximately a half of
$\nu_{R}$ energy. Because two fermions $e^{+}$ and $e^{-}$ has an
isotropic distribution in the $s_{0}$ rest-frame, each electron has a
flat energy spectrum with the maximum energy to be a half of the dark
matter mass. Numerically, we show the energy density distribution in
Fig.~\ref{fig:density}, which is calculated using
Calchep~\cite{Pukhov:2004ca}. 

%
%

As can be seen from Fig.~\ref{fig:density}, the positron energy
density distribution has a flat spectrum with the upper limit to be a
half of the dark matter mass. Since the products of the annihilation
contain mainly leptons, we should anticipate the observation of an
excess in positrons and not in anti-protons~\cite{Adriani:2008zq}. In
order to explain the ATIC-2 results, the dark matter mass $M_{-}$ in
our model should be from 1~TeV to 1.6~TeV. Hence, from
Eq.~(\ref{eq:relic}) the Yukawa coupling $y_{_M}$ needs to be from
$1.2$ to $1.6$ to provide the right relic abundance of dark matter.  

The PAMELA data~\cite{Adriani:2008zr} show a steep increase in the
energy spectrum of the positron fraction $e^{+}/(e^{+}+e^{-})$ in
cosmic rays above 10~GeV. Several groups have analyzed the dark matter
explanation of this observation and found that for dark matter
directly annihilating to two electrons a large boost factor is needed
to fit the PAMELA data~\cite{Cirelli:2008jk}. Depending on diffusion
parameters, a boost factor of  a few hundred  is required in
general~\cite{Cirelli:2008jk}. In our model, we have four electrons in
the final state. The maximum energy for each electron is one half of the
dark matter mass and between 500~GeV  to 800~GeV. Considering the
fact that the electrons have final state radiation of photons, we
anticipate a continuous spectrum with an edge close to $M_{-}/2$. To
explain the PAMELA data, a boost factor from the Sommerfeld enhancement is
needed to fit the observed positron spectrum. In our model, the light
visible particle $s_{0}$ provides a long range force between the dark
matter candidate $\chi_{-}$ and induces a Yukawa potential between two
$\chi_{-}$'s. Neglecting the contact interaction, in the limit $\mu
\ll   y_{_M}^{2}/(4\,\pi)\,M_{-}$, we use the Coulomb potential to
calculate the boost factor due to Sommerfeld
enhancement~\cite{Hisano:2004ds}\cite{ArkaniHamed:2008qn} 
\beq
B\,\approx\,\frac{y_{_M}^{2}}{4\,v_{\rm halo}}\,\approx\,360 \sim 640\,,
\eeq
where $v_{\rm halo}\approx 10^{-3}$ is the typical dark matter
velocity in our Galaxy and $y_{_M}=1.2 - 1.6$ from the relic
abundance calculation. From the analysis in~\cite{Cholis:2008qq}, a
boost factor around 300 for a flat electron energy spectrum with
800~GeV maximum energy provides a good fit to the PAMELA
data. Therefore, up to uncertainties in astrophysical models and
diffusion parameters, our model can accommodate the PAMELA data and at
the same time satisfy the relic abundance.   

\section{DAMA}
\label{sec:dama}
The DAMA collaboration reported an annual modulation in their DAMA/NaI
experiment which has been recently confirmed in the DAMA/LIBRA
experiment by the same collaboration. To reconcile the negative
results from other direct searches such as CDMS, XENON-10 and
CRESST-I, the authors of Ref.~\cite{Bernabei:2007gr} proposed a
scenario in which the dark matter particle interacts dominantly with
the electron in the ordinary matter. In this case, bounds from other
experiments can be avoided. For example, CDMS combines ionization,
phonon and timing information to reject events from electron
recoils. Similarly, XENON rejects electron recoils based on the
ionization/scintillation ratio. In contrast, the DAMA experiments are
based on scintillation only, which can detect electron recoils with a low
threshold. To release energy in the region where the annual
modulation is observed (2-6 keV), elastic scatterings occur between
the dark matter particles and the bound electrons with high momenta
($\sim O(1\, {\rm MeV})$). In NaI (TI), the bound electrons have a small but
non-zero probability to have such high momenta. 

\begin{figure}[ht!]
\centerline{ \hspace*{0.1cm}
\includegraphics[width=0.45\textwidth]{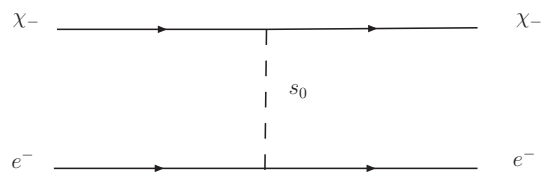}
} 
\caption{Feynman diagram of the DM-electron elastic scattering.}
\label{fig:electron}
\vspace{0.3 cm}
\end{figure}

\begin{figure}[ht!]
\begin{center}
 \includegraphics[width=0.48\textwidth]{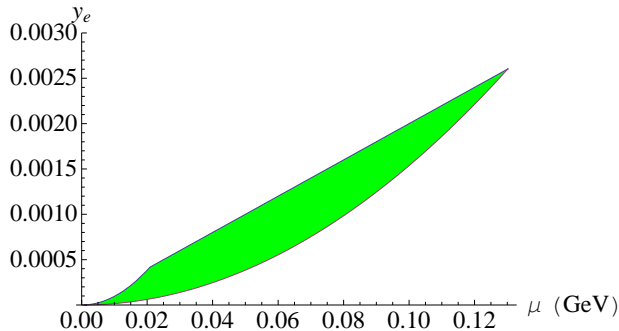}
\caption{\label{fig:dama}The allowed region (shaded) for $\mu$ vs $y_e$.}  
\end{center}
\end{figure}

In our model, the DM-electron scattering is naturally realized by
exchanging the scalar field $s_0$, which is also the mediator to
generate the large boost factor to explain PAMELA. The corresponding
Feynman diagram is shown in Fig.~\ref{fig:electron}.
%
%
In Ref.~\cite{Bernabei:2007gr}, the DAMA/NaI annual modulation data is
analyzed to give a bound 
\begin{equation}
1.1\times10^{-3}\,\mbox{pb}/\GeV<\frac{\xi\,\sigma_e^0}{M_{-}}<42.7\times 10^{-3}\,\mbox{pb}/\GeV\label{eq:dm_e_bound}
\end{equation} 
at $4\sigma$ from the null hypothesis, where $\xi$ is the dark matter
fraction of $\chi^-$ in the halo. In our case $\xi=1$. The
cross-section for DM-electron scattering at rest is denoted
$\sigma_e^0$, and in our model given by
\begin{equation}
\sigma_e^0=\frac{y_e^2\,y_{_M}^2\, m_e^2}{\pi\,\mu^4}.
\end{equation}
The coupling $y_e$ is also constrained by the electron $g-2$: $y_e\lesssim
2\times10^{-5}\,\mu/{\rm MeV}$ (see Appendix \ref{app:g-2}). Assuming $y_{_M}=1.2$ and $M_{-}=1\,{\rm TeV}$,
we obtain the allowed region for $\mu$ and $y_e$ 
from Eq.~(\ref{eq:dm_e_bound}), as shown in Fig.~\ref{fig:dama}.
From Fig.~\ref{fig:dama}, we see that $\mu$ is constrained to be
$\lesssim O(100\,{\rm MeV})$ and the corresponding $y_e$ is consistent with the
fact that it comes from a higher-dimensinal operator. Since the
results from the DAMA/LIBRA experiment confirm the DAMA/NaI results,
we expect a significant allowed region still exists after including the
DAMA/LIBRA data~\cite{DAMAcomment}.

%

\section{Early Annihilation and Diffuse Background}
\label{sec:diffuse}
After dark matter falls out of chemical equilibrium, it may continue
to interact with the standard model fields through elastic
scattering. Therefore, the kinetic equilibrium temperature is in
general below the chemical freeze-out temperature. The existing
studies show that the kinetic decoupling temperature $T_{\rm kd}$ has
a wide range from several MeV to a few GeV in the SUSY and MUED
models~\cite{Profumo:2006bv}. This range of kinetic decoupling
temperatures implies a range of the smallest protohalos with a mass
from $10^{-6} M_{\oplus}$ to $10^{2} M_{\oplus}$.  

\begin{figure}[ht!]
\centerline{ \hspace*{0.1cm}
\includegraphics[width=0.45\textwidth]{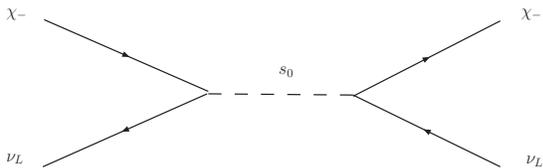}
} 
\caption{Feynman diagram of the elastic scattering of $\chi_{-}$ with $\nu_{L}$.}
\label{fig:neutrino}
\vspace{0.3 cm}
\end{figure}

Specific to our model, if the first scalar KK mode $s_{1}$ has a mass
nearly degenerate with the mass of the dark matter field $\chi_{-}$,
there is an $s$-channel resonance enhancement for the elastic
scattering cross section of $\chi_{-}$ with $\nu_{L}$. Therefore, we
see that a much lower kinetic decoupling temperature $T_{\rm kd}$ can
happen in this model. The relevant Feynman diagram is shown in
Fig.~\ref{fig:neutrino}. 
%
%
When the neutrino energy $E_{\nu}$ is much less than the dark matter
mass, the cross section of this elastic scattering process has the
form 
\beqa
\sigma_{\nu}&=&\frac{y_{_D}^{4}\,E^{2}_{\nu}}{16\,\pi\,[(M_{-}^{2}\,-\,M_{s}^{2})^{2}\,+\,M_{s}^{2}\,\Gamma_{s}^{2}]}
\nonumber \\
&\approx&\frac{y_{_D}^{4}\,E^{2}_{\nu}}{16\,\pi\,(M_{-}^{2}\,-\,M_{s}^{2})^{2}}\,.
\eeqa
For the case $M_{+}>M_{s}>M_{-}$ and $M_{s}-M_{-}< m$, $s_{1}$ decays into $\chi_{-}$ plus $\nu_{L}$ and the width of $s_{1}$ field is 
\beq
\Gamma_{s}\,=\,\frac{y_{_D}^{2}\,(M_{s}^{2}\,-\,M^{2}_{-})^{2}}{16\,\pi\,M_{s}^{3}}\,.
\eeq
For $y_{_D}<1$, we neglect the width part in the propagator of $s_{1}$
and have the cross section only depending on the mass difference of
$s_{1}$ and $\chi_{-}$.  

As the universe expands, the dark matter density and the elastic
scattering rate, $\Gamma_{\nu}\equiv \langle v\,\sigma_{\nu}\rangle
n_{\nu}$, decreases. Here $n_{\nu}$ is the number density of
neutrinos, which are assumed to be in local thermal equilibrium and
$v\approx 1$ in this case. Following the discussion
in~\cite{Green:2005fa}, the thermal average of $\sigma_{\nu}$ is  
\beq
\langle \sigma_{\nu}\,v\rangle\,=\,\frac{9\,y_{_D}^{4}\,T^{2}}{64\,\pi\,(M_{-}^{2}\,-\,M_{s}^{2})^{2}}\,.
\eeq
As functions of temperature, $n_{\nu}\sim T^{3}$ and the Hubble rate
of expansion $H\sim T^{2}/m_{\rm pl}$. The relaxation time $\tau$
is defined as the time $\chi_{-}$'s need to return to local thermal
equilibrium after a deviation from it, which is related to the
elastic scattering rate as $\tau(T)\approx
\sqrt{2/3}\,M_{-}/(T\,\Gamma_{\nu})$. The kinetic decoupling
of the dark matter candidate $\chi_{-}$ happens when
$\tau(T_{\rm kd})\,=\,1/H(T_{\rm kd})$, from which we obtain 
\beq
T_{\rm kd}\,\approx\,\frac{2}{y_{_D}}\,\left(\frac{M_{-}}{m_{\rm pl}}\right)^{1/4}\,\Delta\,,
\eeq
where $\Delta^{2}\equiv M_{-}^{2}\,-\,M_{s}^{2}$. For example, when
$M_{-}=1.0$~TeV and $y_{_D}=0.1$, $T_{\rm kd}$ varies from 2~keV to
20~MeV for $\Delta$ between 1~MeV and 10~GeV. Using the relation
between the mass of the first gravitational-bound structure, $M_{c}$, and the
kinetic decoupling temperature \cite{Loeb:2005pm}: 
\beq
M_{c}\simeq 33\,(T_{\rm kd}/10~{\rm MeV})^{-3}\,M_{\oplus}\,,
\eeq
we have $300\,M_{\oplus}< M_{c} < 3\times
10^{14}\,M_{\oplus}$ for $\Delta$ between  10~GeV and 1~MeV.  

The $\chi_{-}$'s in the dark-matter halos annihilate into
electron-positron pairs in the energy of a few hundred GeV. The
electrons and positrons rapidly inverse-Compton scatter with CMB
photons and contribute to the diffuse extragalactic gamma-ray
background today. The energy density in photons today from dark matter
annihilation in the first halos is calculated
in~\cite{Kamionkowski:2008gj} as
\beq
\rho_{\gamma}\,\approx\,2.64\times 10^{-11}\,\left(\frac{M_{c}}{M_{\oplus}}\right)^{-1/3}\,\left( \frac{M_-}{\rm TeV}  \right)^{-1}~{\rm GeV\,cm^{-3}}\,.
\eeq
The EGRET experiment imposes a bound on the extragalactic gamma-ray
background~\cite{Sreekumar:1997un}. It can be translated to
$\rho_{\gamma}\leq 5.7\times 10^{-16} (E_{\gamma}/{\rm
  GeV})^{-0.1}~{\rm GeV\,cm^{-3}}$. Therefore, this imposes a bound on
$\Delta$, which is the mass square difference between $\chi_{-}$ and
$s_{1}$, as 
\beqa
\Delta
\leq\frac{y_{_D}}{0.5}\,\left( \frac{M_{-}}{1\,{\rm TeV}}
\right)^{3/4}\,\left(\frac{E_{\gamma}}{{\rm
    GeV}}\right)^{-0.1}\,1.2~{\rm MeV}\,. 
\eeqa
The access energy range of $E_{\gamma}$ in EGRET is from $30$~MeV to
$100$~GeV. This means that a degenerate spectrum between $\chi_{-}$
and $s_{1}$ up to order of MeV is needed to evade the current bound from
the diffuse background.  

\section{Discussions and Conclusions}
\label{sec:conclusion}

At the LHC, the production mechanism of the KK-odd particles in our
model is similar to the minimal UED model. Unlike the minimal UED
model,  the KK mode of the right-handed neutrino $\chi_{-}$ is the
lightest KK-odd particle. Hence all other first KK modes of the SM
particles should ultimately decay into $\chi_{-}$. Interestingly,
the KK photon $B^{1}$, which is the lightest KK-odd particle in the
minimal UED, decays into $s_{1}$ plus two electrons through an
off-shell intermediate KK electron $e^1$ exchanging. The $s_{1}$
subsequently decays into $\chi_{-}$ and $\nu_{L}$. The decay
process of $B^1$ is: 
\beqa
B^1\, \xrightarrow{e^1}\,e^+\,+\,e^-\,+\,s_1\,\rightarrow\, e^+\,+\,e^-\, +\,\chi_-\,+\nu_L\,.
\eeqa
If the mass difference between $B^{1}$ and $\chi_{-}$ is a few tens of
GeV or more, there will be lots of energetic leptons produced at the
LHC~\cite{ArkaniHamed:2008qp}. In order to accommodate the DAMA results
and be consistent with the electron $g-2$, the relevant coupling
$y_{e}$ is of order $10^{-3}$. Hence the width of $B^{1}$
is estimated to be $\sim y_{e}^{2}\,e^{2}\,\Delta M/(64\,\pi^{3})$ with  $\Delta
M^{2}=M^{2}_{B^{1}}-M_{-}^{2}$, which is of order eV. 

Since the products of the dark matter annihilation also contain
high energy neutrinos, the Super-Kamiokande may observe those energetic neutrinos
from the sun~\cite{Desai:2004pq}. When dark 
matter meets the sun, its speed will be slowed down due to its elastic
scattering with electrons in the sun. Once the dark matter speed is
reduced below the gravitational escaping velocity, it will be
captured by the sun and produce additional neutrinos through
annihilation. We leave this neutrinos flux calculation related to
Super-Kamiokande to future study. 

In conclusion, we have explored the sUED model, which is an extension
of the UED model by including SM singlets, to explain the overwhelming
evidence of direct and indirect dark matter detections from
experiments including DAMA, ATIC-2, PPB-BETS and PAMELA. The dark
matter candidate is the first KK-mode of the right-handed neutrino,
$\chi_{-}$, whose stability is protected by the KK-parity.  

The dark matter candidate $\chi_{-}$ mainly annihilates into the
right-handed neutrino, which subsequently decays into the left-handed
neutrino and a light SM singlet scalar, $s_{0}$. The scalar $s_{0}$ has
a mass below 1 GeV, which mainly decays into two electrons. Therefore,
the final state particles of dark matter annihilation contain four
electrons and two neutrinos. To explain the electron and positron
energy spectrum observed by ATIC-2 and PPB-BETS, we found that the
mass of the dark matter candidate should be from 1~TeV to 1.6~TeV,
which sets the fifth dimension compactification scale. The PAMELA
result is explained by the same dark matter annihilation. The needed
``boost factor'' in the cross-section is obtained through the
Sommerfeld enhancement effect, due to the long-range force between two
dark matter particles by exchanging 
$s_{0}$. The dark matter relic abundance determines the value of the
Yukawa coupling of the dark matter to the 
scalar singlet. The same Yukawa coupling determines the boost factor
from the Sommerfeld effect to be $360-640$, suitable for explaining
the PAMELA resuts.  

The DAMA results are explained by the elastic scattering of $\chi_{-}$
with electrons through exchanging the light scalar field $s_{0}$ in
the $t$-channel. Since $s_{0}$ only couples to leptons, the null
results of the dark matter direct searches at CDMS and XENON, which
veto electron recoils, are automatically explained.  We have found
that there exists parameter space in our model to accommodate the DAMA
results without contradicting the electron $g-2$. Finally, by
calculating the $s$-channel elastic scattering cross section of
$\chi_{-}$ with the left-handed neutrino by exchanging the first KK
mode of the scalar field $s_{1}$, we show that the diffuse
extragalactic gamma-ray background constrains can be satisfied
provided that the masses of $\chi_{-}$ and $s_{1}$ are highly
degenerate.

\bigskip

{\bf Acknowledgments:} 
Many thanks to Patrick Fox  for interesting discussions and Marco
Cirelli for useful correspondences.   Z.H. is supported  in part by
the United States Department of Energy grand
no. DE-FG03-91ER40674. Fermilab is operated by Fermi Research
Alliance, LLC under contract no.  DE-AC02-07CH11359 with the United
States Department of Energy.   

\begin{appendix}

\section{The constraint to $y_e$ from electron $g-2$.}
\label{app:g-2}
The current experimental value for electron $g-2$ is given by \cite{g-2}
\begin{equation}
a_e=(1\, 159\, 652\, 180.85\pm.76)\times10^{-12}\,.
\end{equation}
Given uncertainties in the determination of $\alpha$, extra
contributions to $a_e$ should satisfy \cite{Fayet:2007ua}
\begin{equation}
|\delta a_e|\lesssim2\times 10^{-11}\,.\label{eq:deltaa_bound}
\end{equation}
From the triangle diagram of $s_0$ exchange, we have \cite{Krawczyk:1996sm}
\begin{equation}
\delta a_e = \frac{y_e^2}{8\,\pi^2}\,\widetilde{L}\,,\label{eq:deltaa}
\end{equation}
where
\begin{equation}
\widetilde{L}=\int_0^1\mbox{d} x\,\frac{x^2(2-x)}{x^2+(1-x)(\mu/m_e)^2}\,.\label{eq:L}
\end{equation}
When $\mu\gg m_e$, Eqs.~(\ref{eq:deltaa_bound}), (\ref{eq:deltaa}) and (\ref{eq:L}) give us 
\begin{equation}
y_e\lesssim 2\times 10^{-5}\frac{\mu}{\rm MeV}\,. 
\end{equation}
\end{appendix}

 
 \end{document}